\documentclass[sigconf]{acmart}
\usepackage{courier}  
\usepackage{url}            
\usepackage{amsfonts}       
\usepackage{nicefrac}       
\usepackage{xcolor}         
\usepackage{mathrsfs}
\usepackage{microtype}
\usepackage{multirow}
\usepackage{booktabs}

\usepackage{amsmath, amssymb}
\usepackage{mathrsfs}
\usepackage{subfigure}
\usepackage{pgf-pie}
\usepackage{pgfplots}
\pgfplotsset{compat=1.18}
\usepackage{natbib}  
\usepackage{caption} 
\usepackage{algorithm}
\usepackage{algorithmic}

\usepackage{newfloat}
\usepackage{listings}
\AtBeginDocument{%
  \providecommand\BibTeX{{%
    \normalfont B\kern-0.5em{\scshape i\kern-0.25em b}\kern-0.8em\TeX}}}

\setcopyright{acmcopyright}
\copyrightyear{2023}
\acmYear{2023}
\setcopyright{acmlicensed}\acmConference[MM '23]{Proceedings of the 31st
ACM International Conference on Multimedia}{October 29-November 3,
2023}{Ottawa, ON, Canada}
\acmBooktitle{Proceedings of the 31st ACM International Conference on
Multimedia (MM '23), October 29-November 3, 2023, Ottawa, ON, Canada}
\acmPrice{15.00}
\acmDOI{10.1145/3581783.3613800}
\acmISBN{979-8-4007-0108-5/23/10}

\begin{document}

\title{PMVC: Data Augmentation-Based Prosody Modeling for Expressive Voice Conversion}

\author{Yimin Deng}
\authornote{These authors contributed equally to this research.}
\affiliation{
    \institution{Ping An Technology (Shenzhen) Co., Ltd.}
    \institution{University of Science and Technology of China}
    \country{China}
    }
\email{dengyimin0312@mail.ustc.edu.cn}

\author{Huaizhen Tang}
\authornotemark[1]
\affiliation{%
\institution{Huya Inc (Shenzhen) Co., Ltd.}
  \institution{Ping An Technology (Shenzhen) Co., Ltd.}
  \country{China}
}

\email{tanghuaizhen@huya.com}

\author{Xulong Zhang}
\authornotemark[1]
\affiliation{%
  \institution{Ping An Technology (Shenzhen) Co., Ltd.}
  \country{China}
}
\email{zhangxulong@ieee.org}

\author{Jianzong Wang}
\authornote{Corresponding author: Jianzong Wang (jzwang@188.com).
}
\affiliation{%
  \institution{Ping An Technology (Shenzhen) Co., Ltd.}
  \country{China}
}
\email{jzwang@188.com}

\author{Ning Cheng}
\affiliation{%
  \institution{Ping An Technology (Shenzhen) Co., Ltd.}
  \country{China}
}
\email{chengning211@pingan.com.cn}

\author{Jing Xiao}
\affiliation{%
  \institution{Ping An Technology (Shenzhen) Co., Ltd.}
  \country{China}
}
\email{xiaojing661@pingan.com.cn}

\renewcommand{\shortauthors}{Yimin Deng et al.}
\begin{abstract}
Voice conversion as the style transfer task applied to speech, refers to converting one person's speech into a new speech that sounds like another person's. Up to now, there has been a lot of research devoted to better implementation of VC tasks. However, a good voice conversion model should not only match the timbre information of the target speaker, but also expressive information such as prosody, pace, pause, etc. In this context, prosody modeling is crucial for achieving expressive voice conversion that sounds natural and convincing. Unfortunately, prosody modeling is important but challenging, especially without text transcriptions. In this paper, we firstly propose a novel voice conversion framework named `PMVC', which effectively separates and models the content, timbre, and prosodic information from the speech without text transcriptions. Specially, we introduce a new speech augmentation algorithm for robust prosody extraction. And building upon this, mask and predict mechanism is applied in the disentanglement of prosody and content information. The experimental results on the AIShell-3 corpus supports our improvement of naturalness and similarity of converted speech.

\end{abstract}





\maketitle
\section{Introduction}
Voice Conversion(VC), also called voice style transfer, aims to modify the voice characteristic of the source speech to convert one speaker's voice to generate a new target speech as it is said by another people. It covers a wide range of applications like intelligent security products and aids. So far, many algorithms have made much progress in VC successfully~\cite{GMM,GAN,VAE,vq-wav2vec,EVC}. One of the most popular methods is to achieve VC tasks by separating the content and timbre information of speech to learn the disentangled speech representations~\cite{NVC-net, vq-vae, tgavc, vq-wav2vec}. Specifically, these algorithms usually follow the autoencoder framework, and the encoder is trained to learn to represent content information and the timbre information, respectively. At the same time, a decoder is trained to output a natural speech from given content and timbre representations. With the pretrained autoencoder, we only need to replace the timbre representations of the source speech with that of the target speech before fed into decoder to generate the converted result.

Although these algorithms can easily implement VC tasks, few of them can convert all voice characteristics as expected~\cite{attention-wavenet}, such as the prosodic information. In fact, with previous algorithms, we can observe that almost all converted speech have the same pace, pause, and pitch contour shape as the source speech no matter what these voice characteristics of the target speech. In other words, previous methods convert only timbre, not all voice characteristics.
Nowadays, VC not only demands high naturalness but also requires sufficient expressiveness in various scenarios, such as automatic movie dubbing with emotional conversations. Hence, with the advancement of deep learning, prosody modeling for expressive voice conversion has gradually attracted more and more attention~\cite{M2VOC,icml3, speechsplit,autopst}. Its introduction greatly enriches the diversity and expressiveness of converted speech. Integrating prosody modeling into VC models provides basic framework for fine-tunning to carry out downstream tasks like emotional speech processing~\cite{tang2023qi,tang2023EmoMix}. 
Specifically, a speech can be roughly decomposed into three components: the content information, which characterizes the phoneme and linguistic information; The timbre information, which is closely related to the speaker identity; Besides, the pace, pause and rhythm of a speech, which we call all of them the prosodic information. Obviously, in order to improve the expressiveness of voice conversion, compared with the simple division of speaker-independent and speaker-related information, we need to further model the content, timbre, and prosodic information, respectively. 

In this paper, we introduce data augmented-based Prosody Modeling Voice Conversion (PMVC) model, which performs expressive voice conversion based on a novel speech augmentation algorithm. PMVC utilizes adaptive instance normalization in its encoder to eliminate the global static information from the speech. In addition, some information-theory-guided approaches are used to model the disentangled linguistic and prosodic features efficiently. Unlike many other previous models, PMVC models the prosody features without any text transcriptions. It significantly simplifies the complexity of the model and allows the prosody representation to contain richer information about the expressiveness of the phoneme rather than just the phoneme duration. 

\section{Related Work}
\subsection{Voice Conversion}
Voice conversion (VC) is a task that aims to transfer speaking style of input speech while preserving content information. GAN-based methods~\cite{cyclegan,stargan-vc} are employed to perform VC task with advantages of high efficiency and diversity. Speaking style is regarded as a condition and injected into generation process. However, such training process lacks of controllability. 

Hence, disentanglement based VC methods have aroused people's attention. Qian \textit{et. al.} first proposed AutoVC~\cite{autovc}, by applying the bottleneck structure, AutoVC can force the encoder to discard some information of the input speech to learn the disentangled content representation. Vector quantization~(VQ) based methods~\cite{avqvc,tang2023vqcl} and adversarial training based methods~\cite{tang2023clsvc} are also introduced for better disentanglement. Then Studies have been conducted to explore more voice characteristics for expressiveness, such as prosody. 

\subsection{Prosody Modeling}
Proosdy modeling is crucial yet challenging. The causes of this complexity are multifaceted, and one critical reason is the difficulty in completely eliminating prosodic information from the source speech. As~\cite{emo_style} said, prosodic features are often closely associated with phonemes, so it is difficult to separate them and model them individually. Currently, prosody modeling methods are predominantly present in expressive text-to-speech (TTS) systems. Previous work~\cite{icml1} introduced a Tacotron-based method capable of disentangling prosody from speech content.
Mellotron~\cite{Mellotron} further extracts different aspects of the prosodic information. Besides, CHiVE~\cite{ICML2} is also proposed to extract and learn prosody features for expressive TTS. Recently, prosody modeling is also applied to VC tasks. Parrotron~\cite{Parrotron} extract prosodic information by encouraging the latent code to be the same as the phoneme embeddings, IQDUBBING~\cite{iqdubbing} use a prosody extracter and two prosody filter to extract the prosodic features. However, all of the above methods require text transcriptions. Of course, with text transcriptions, prosody modeling can be easier, but it also limits their ability to scale to those speech corpus that don't have text transcriptions. 

To further improve the expressiveness of VC, SpeechFlow~\cite{speechsplit} achieves the disentanglement of the voice characteristics like content, timbre, rhythm and pitch information from the input speech. Besides, by defining prosodic information as the duration of a phoneme, AutoPST~\cite{autopst} achieves the global rhythm transfer without any text transcriptions. Both SpeechFlow and AutoPST necessitate imposing stringent constraints on the dimensions of the latent embedding to achieve a proper balance between timbre disentanglement and intelligibility. It's related to the final quality of conversion and poses challenges for direct application to other datasets.

To issue this problem, in this paper, we first discard the bottleneck structure. At the same time, we introduce the Instance Normalization layer to achieve a similar function of filtering timbre information. In fact, previous studies have shown that Instance normalization(IN) can eliminate the speaker information without introducing any information bottleneck structure~\cite{IN,IN1}. Recently, INVC~\cite{INVC} separated the speaker and content representations by applying the adaptive instance normalization. However, as we said before, IN just eliminates the global timbre information, and the time-variant representations still contain the phoneme and prosodic information. We still need an effective method to extract the prosodic features from the content representations.

Inspired by~\cite{avqvc}, we proposed a new method that utilizes contrastive learning to disentangle the content embeddings and the prosody embeddings from the time-variant representations. In order to implement this method, we need to first construct the augmented speech of the original speech. 

\begin{figure*}[t]
    \subfigure[Framework of PMVC]{
        \label{fig:1(a)}
        \begin{minipage}[b]{0.66\linewidth}
            \centering
        \includegraphics[width=0.9\linewidth]{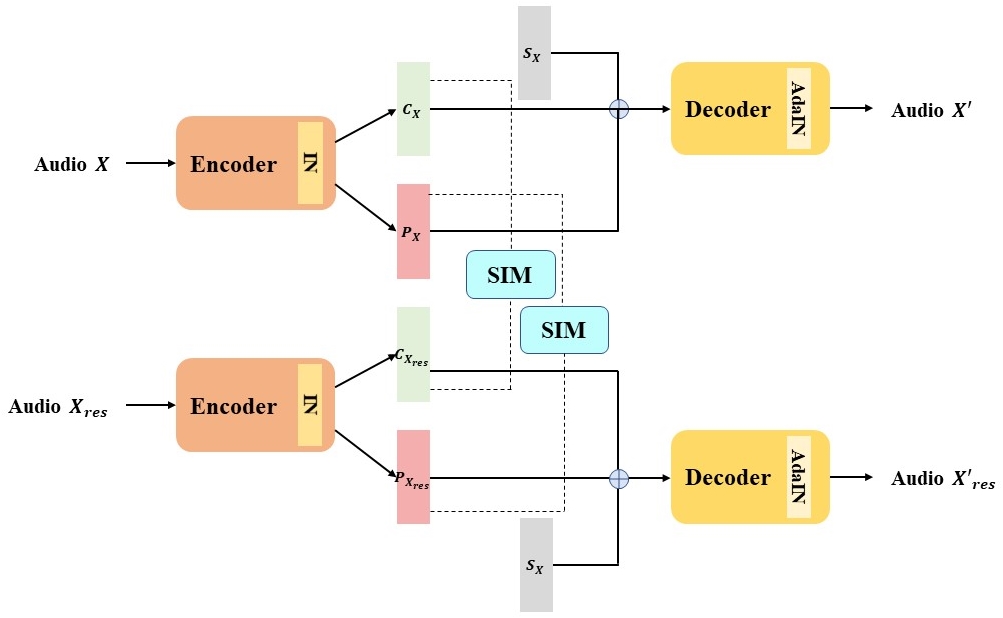}
        \end{minipage}
    }
    \subfigure[Content predictor]{
        \label{fig:1(b)}
        \begin{minipage}[b]{0.3\linewidth}
            \centering
        \includegraphics[width=0.7\linewidth]{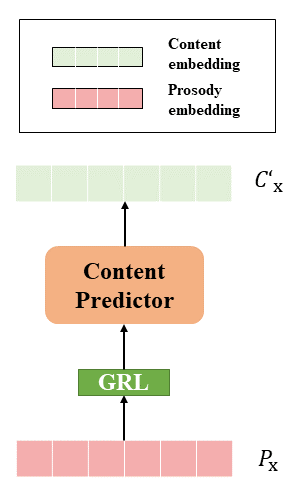}
        \end{minipage}
    }
    
    \caption{Framework of PMVC. $\boldsymbol C_x$ and $\boldsymbol P_x$ are the content features and prosodic features extracted from the input speech. Denote $\boldsymbol S_x$ as speaker embedding, generated from the pretrained speaker encoder. $\boldsymbol{IN}$, $\boldsymbol{AdaIN}$ stand for Instance Normalization and adaptive instance normalization, respectively. Which can eliminate the global static information from $\boldsymbol{x}$. The right image shows the content predictor. $\boldsymbol C'_x$ denotes the predicted content embedding, it’s reasonable to expect a close association with $\boldsymbol C_x$. \textbf{GRL} means Gradient Reversal Layer, it will make the optimization goal of the feature encoder and the content predictor completely opposite.}
\end{figure*}

\section{Proposed Methods}
As depicted in Figure~\ref{fig:1(a)}, the framework of PMVC includes three main modules. The first one is a feature encoder $E$, responsible for the extraction of content feature $\boldsymbol{C}$ and the prosody feature $\boldsymbol{P}$ from the input speech $\boldsymbol{X}$. $IN$ in $E$ means Instance Normalization, and it can remove the global static information from $\boldsymbol{X}$. Instead, a pretrained speaker encoder $E_s$ is introduced to provide the speaker embedding $\boldsymbol{S}$. With the disentangled content embedding, prosody embedding, and the speaker embedding, a decoder $D$ is trained to output a natural reconstructed speech $\boldsymbol{X'}$.
\subsection{Stretching Audio Time Series Strategy}
In this paper, the spectrogram of the original audio time series is denoted as $X(T)$, where $T$ represents the frame number. Then, we define the content vector $C$ to represent the content information, the prosody vector $P$ to represent the prosodic information, and we define the timbre vector $S$ as the speaker-related information. Based on our assumption, for each speech segment $X$, it can be uniquely determined by given the disentangled speech representations $C$, $P$ and $S$. Formally, $X$ can be considered a random variable sampled from the speech distribution $p_x(\cdot | C, P, S)$.  

As mentioned in SpeechFlow~\cite{speechsplit}, Random Resampling~(RR) operation is an effective strategy to change the prosodic information of the original audio time series. Specifically, RR involves three steps of operations. The initial step involves dividing the input audio series into segments of random lengths. Then, the second step is to randomly draw a sampling rate for each segment. And, the last step is to re-sample the segment with the selected sampling rate. Compared with the original audio sequence $X$, the output audio sequence $X_{RR}$ retains the original content, but changes the timbre and prosodic information. It can be expressed as:
\begin{align}
    \nonumber\boldsymbol{X} &\sim p_X(\boldsymbol{\cdot | C = C_X, P = P_X, S = S_X}).\\
    \boldsymbol{X_{RR}} &\sim p_{X_{RR}}(\boldsymbol{\cdot | C = C_X, P = P_{X_{RR}}, S = S_{X_{RR}}} )
\end{align}

Noted that after random resampling, the sequence length of the audio may also be changed. To address this problem, existing methods align the lengths of $X$ and $X_{RR}$ by using a padding constant of 0~\cite{speechsplit, RR_align}. However, it will inevitably affect the data quality, thereby increasing the difficulty of model training. Besides, in order to make prosodic modeling more convenient, we expect to find a new algorithm that can change the prosodic information but remain the original timbre and content, that is, Time scale modification (TSM).

\begin{algorithm}[tb]
\caption{Random Prosody Algorithm}
\label{alg:algorithm}
\begin{flushleft}
\textbf{Input}: A speech segment $\boldsymbol{X}$ of length $T$\\
\textbf{Parameter}: The split length $\boldsymbol{t}$, sampling rate $R$\\
\textbf{Output}: RR speech segment 
$\boldsymbol{X_{res}}$ with same length $T$\\
\end{flushleft}
\begin{algorithmic}[1] 
\STATE Let $num=T/t$. We divide $\boldsymbol{X}$ into $num$ segments with the same length $\boldsymbol{t}$, $L = [x_1, x_2, ... x_{num}]$  
\WHILE{$num > 1$}
\STATE Select two segments $\boldsymbol{x_i}$ and $\boldsymbol{x_j}$ from $L$
\STATE $a = Uniform(0.6, 2)$
\STATE Stretch audio time series $\boldsymbol{x_i}$ with the specific rate $a$
\STATE Stretch audio time series $\boldsymbol{x_j}$ with the specific rate $\frac{a}{2a-1}$
\STATE Remove $\boldsymbol{x_i}$ and $\boldsymbol{x_j}$ from $L$
\STATE $num = num-2$
\ENDWHILE
\STATE $\boldsymbol{X_{res}}$ is obtained by concatenating all speech segments in the original order
\STATE \textbf{return} $\boldsymbol{X_{res}}$
\end{algorithmic}
\end{algorithm}

Inspired by it, this paper proposes a new method that guides the learning of disentanglement speech representations with information-theory-guided constraints. Specifically, we first propose a new strategy of stretching audio time series, which can be roughly divided into three steps. As shown in~\textbf{Algorithm \ref{alg:algorithm}}, first, the input sequence is segmented into uniform-length segments. Second, for each pair of segments selected randomly, a rate is randomly drawn, and the total sample points of the sequence segments remain the same (one segment is stretched, and the other is correspondingly shortened). Finally, we stitch all the speech fragments together in the previous order. With this algorithm, for each sequence $x$, we can get a corresponding augmented speech $x_{res}$ with the same length $T$. And it can be expressed as:
\begin{align}
    \boldsymbol{X_{res}} \sim p_{X_{res}}(\boldsymbol{\cdot | C = C_X, P = P_{X_{res}}, S = S_X})
\end{align}
And based on the augmented data, a new method has been proposed to extract the prosodic information from speech.

\subsection{How to Train The Model}

Here we will present how and why our model can induce the content embedding and prosody embedding into independent representation spaces simultaneously.

As we discussed before, $\boldsymbol{X_{res}}$ can be regarded as an augmented speech of $\boldsymbol{X}$. That is, $\boldsymbol{X_{res}}$ and $\boldsymbol{X}$ have the same content information, same timbre information and different prosodic information. During training, a pair of speech segments ($\boldsymbol{x, x_{res}}$) are selected to be the input, the feature encoder $E$ can eliminate the global speaker information while preserving other information from the input speech by using instance normalization without affine transformation, and it can be expressed as:
\begin{align}
    &\mu_c = \frac{1}{W}\sum_{w=1}^WM_c[w]\\
    &\alpha_c = \sqrt{\frac{1}{W}\sum_{w=1}^W(M_c[w]-\mu_c)^2+\epsilon}\\
    &Z_c[w] = \frac{M_c[w]-\mu_c}{\alpha_c}
\end{align}

\noindent where $M_c$ is the feature map in $c$-th channel, $W$ denote as the dimension of $M_c$, $M_c[w]$ is the $w$-th element in $M_c$, $Z_c[w]$ refer to the normalized $M_c[w]$. Besides, $\epsilon$ is a small value which can avoid numerical instability. 

Obviously, the normalized hidden feature $Z$ contains both content information and prosodic information. We further hypothesize that $Z$ is a specific expression composed of the estimated content embedding $\boldsymbol{C_x}$ and estimated prosody embedding $\boldsymbol{P_x}$:
\begin{align}
   \boldsymbol Z = E(\boldsymbol{x}) = \boldsymbol{C_x} \oplus \boldsymbol{P_x}
\end{align}
where $\oplus$ means concatenation. In this scenario, we split $\boldsymbol{Z}$ along the channel dimensions, representing the estimated content embedding $\boldsymbol{C_x}$ and the estimated prosody embedding $\boldsymbol{P_x}$ respectively.

Since $\boldsymbol{x}$ and $\boldsymbol{x_{res}}$ have the same content information, their content embeddings are expect to be as closer as possible. At the same time, after Random Prosody (RP) operation, the prosodic information of $\boldsymbol{x}$ have been corrupted. In other words, the prosodic information in $\boldsymbol{x}$ and $\boldsymbol{x_{res}}$ should be different. Hence, we expect their prosody embeddings should be as different as possible. However, although $\boldsymbol{x}$ and $\boldsymbol{x_{res}}$ contain the same semantic information, the phoneme of each frame may be different (otherwise $\boldsymbol{x}$ and $\boldsymbol{x_{res}}$ will be exactly the same). So, unlike most similar studies, MSE loss or L1 loss cannot be applied here. Specially, we employ cosine similarity to measure the similarity between a pair of features in this paper:
\begin{align}
    G(A(x), A(x_{res})) = \frac{A^T(x)A(x_{res})}{\|A(x)\|_2\|A(x_{res})\|_2}
\end{align}

\noindent where $G(\cdot, \cdot)$ represents the calculation of cosine similarity score. $A(\cdot)$ can be used to represent any extracted embedding of input speech.

As mentioned, the training of the proposed model aims to maximize cosine similarity between similar content embeddings, and minimize it between the different prosody embeddings. Hence, the proposed similarity contrastive loss function for model training is:
\begin{align}
   \mathcal{L_\text{sim}} = \frac{G(P(x), P(x_{res}))}{G(C(x), C(x_{res}))}
\end{align}

Besides, the speaker embedding $S_x$ is produced by a pretrained speaker encoder with GE2E loss~\cite{GE2E}. It involves positive pairs composed of different utterances of the same speaker and negative pairs composed of different speakers. The embedding similarity of positive pairs needs to be maximized, and the similarity of negative pairs needs to be minimized during pretraining. We can easily find that the speaker embedding contains only the timbre information. Here we will give a formal discussion about the speaker embedding:
We assume that there are two speakers $S1$ and $S2$ and some of their speeches. As we discussed before, for each speech $X$ belong to speaker $S$, it can be formulated as:
\begin{equation}\nonumber
   \begin{aligned}
    \nonumber\boldsymbol{X} &\sim p_X(\boldsymbol{\cdot | C = C, P = P, S = S}).\\
    \end{aligned} 
\end{equation}

Now, assume there are two speeches $x1$ and $x2$, both from the same speaker $S1$. And, there are another speech $x3$ belongs to another speaker $S2$. It can be expressed as:
\begin{equation}\nonumber
    \begin{aligned}
    \nonumber\boldsymbol{x1} &\sim p_X(\boldsymbol{\cdot | C = C_{x1}, P = P_{x1}, S = S1}).\\
    \nonumber\boldsymbol{x2} &\sim p_X(\boldsymbol{\cdot | C = C_{x2}, P = P_{x2}, S = S1}).\\
    \nonumber\boldsymbol{x3} &\sim p_X(\boldsymbol{\cdot | C = C_{x3}, P = P_{x3}, S = S2}).\\
\end{aligned}
\end{equation}

Note that the content and prosody information of each speech here are random. For the convenience of discussion, we assume that the content information of $x1$ and $x3$ are the same. In other words, the only difference between $x1$ and $x3$ are the timbre information and part of prosody information. In training, we expect the speaker encoder can output different embeddings from $x1$ and $x3$. Then, the most convenient way for the speaker encoder in training is to discard the content information and extract the timbre and prosodic features. 

At the same time, we can further assume that the prosody information in $x1$ and $x2$ are different. In training, we expect the speaker encoder would output the same embeddings from $x1$ and $x2$. In this case, the encoder would be encouraged to discard the prosody information and extract the timbre and part of the content features. 

In summary, the speaker encoder will learn to extract only the timbre information in the speech, while eliminating the content and prosody information as much as possible. So we say the speaker embeddings contain only the timbre information.

Finally, leveraging the content embedding $C_x$, prosody embedding $P_x$, and speaker embedding $S_x$, the decoder is guided to produce the reconstructed speech $x'$. We employ a reconstruction loss function during training, which is as follows:
\begin{align}
    \mathcal{L}_{\text {recon }} &= \|x'-x\|_{2}^{2} +  \|x'_{res}-x_{res}\|_{2}^{2}
\end{align}

\noindent where $x'$ is generated from $C_x$, $P_x$ and $S_x$, $X'_{res}$ is produced from $C_{x_{res}}$, $P_{x_{res}}$ and $S_x$.

\subsection{Mask and Predict}

In the above subsection, we introduced the similarity contrast loss $\mathcal{L}_{\text {sim}}$ and the reconstruction loss $\mathcal{L}_{\text {recon }}$ to encourage our model to learn the disentangled speech representations. But, considering such a case, with only the above two loss function constraints, the most convenient way for the proposed model is to copy both the content and prosodic information of $x$ to $P_x$, while the content embedding $C_x$ contains no information. In this case, $C_x = C_{x_{res}} = 0$, and $\mathcal{L}_{\text {sim}}$ will be optimized to zero. At the same time, since $P_x$ and $S_x$ are able to provide all the information needed for speech reconstruction, $\mathcal{L}_{\text {recon }}$ can also be optimized to zero. In other words, the above two objective functions can not prevent this special case from appearing. However, in the inference phase, it will make the target content information leak into the decoder, which will lead to failed VC tasks.

To issue this problem, we need to force the prosody embedding contains no content information. Inspired by Mask-Predict~\cite{MP}, we proposed an adversarial training way to remove some information from the prosody embedding. Specifically, a Gradient Reversal layer (GRL)~\cite{GRL} between the feature encoder and a content predictor is introduced. When we put the hidden feature $Z$ into the content predictor, we first mask the first part of $Z$. That is, only the information contained in the estimated prosody embedding will be used to predict the estimated content embedding. As illustrated in Figure~\ref{fig:1(b)}, during training, we put the prosody embedding $P_{x}$ into the content predictor, and the content predictor would be expected to output the content embedding as accurately as possible. At the same time, due to the GRL, the feature encoder and the content predictor have opposite optimization goals. In other words, the feature encoder would be encouraged to eliminate the content information contained in the estimated prosody embeddings. Finally, the prosody embedding will remove some information so that the content predictor can not reconstruct the masked estimated content embedding. The adversarial loss can be formulated as:
\begin{align}
    \mathcal{L}_{\text {adv}} &= \|C'_x-C_x\|_{2}^{2} +  \|C'_{x_{res}}-C_{x_{res}}\|_{2}^{2}
\end{align}

\noindent where $C'_x$ is generated from $P_x$, $C'_{x_{res}}$ is produced from $P_{x_{res}}$. Here we use $\theta_e$ and $\theta_p$ to respectively represent the trainable parameters of the feature encoder and the content predictor. We use \textit{Pred} to represent content predictor, then the optimization goal is 
\begin{align}
     E^{*}, Pred^{*}= \arg \min_{\theta_p} \max_{\theta_e} \mathcal{L}_{adv}
\end{align}

As we said before, the content predictor is optimized to minimize $ \mathcal{L}_{\text {adv}}$, while the feature encoder is optimized to maximized $ \mathcal{L}_{\text {adv}}$. As a result, this loss function will converge when $P_x$ discards some information contained in the estimated content embeddings $C_x$. In addition, the feature encoder is also optimized to minimize $ \mathcal{L}_{\text {sim}}$ and $ \mathcal{L}_{\text {recon}}$. As a result, $P_x$ will be encouraged to remove the same content information but preserve the different prosodic information to minimize $ \mathcal{L}_{\text {sim}}$. Furthermore, to minimize $ \mathcal{L}_{\text {adv}}$ and $ \mathcal{L}_{\text {recon}}$ , the estimated content embedding $C_x$ would be encouraged to carry all content information to achieve a well speech reconstructed task.


The complete loss function can be a combination of weighted loss items mentioned above as follows:

\begin{align}
    \label{eq:9}
    L(\boldsymbol{\theta_e, \theta_d}) = \mathcal{L}_{\text {recon}} + \alpha \mathcal{L}_{\text {sim}} + \beta \mathcal{L}_{\text {adv}}
\end{align}

\noindent where $\theta_e$ and $\theta_d$ indicate the trainable parameters of the feature encoder and the decoder, respectively. $\alpha$ , $\beta$ are hyper-parameters as the weight of $\mathcal{L}_{\text {sim}}$ and $\mathcal{L}_{\text {adv}}$, respectively.

Now we can say, with the full loss function $ L(\boldsymbol{\theta_e, \theta_d})$. Our model will be trained to learn the disentangled speech representations for expressive voice conversion.

\subsection{Architecture of the Proposed Framework}

\begin{figure}[t]
    \subfigure[Feature encoder]{
        \label{fig:3(a)}
        \begin{minipage}[b]{0.45\linewidth}
            \centering
    \includegraphics[width=0.8\linewidth]{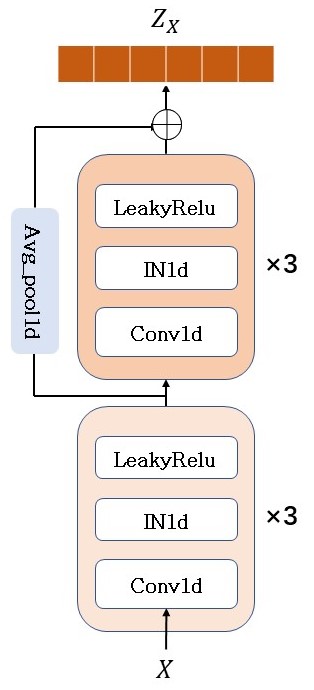}
        \end{minipage}
    }
    \subfigure[Content predictor]{
        \label{fig:3(b)}
        \begin{minipage}[b]{0.45\linewidth}
            \centering
    \includegraphics[width=0.8\linewidth]{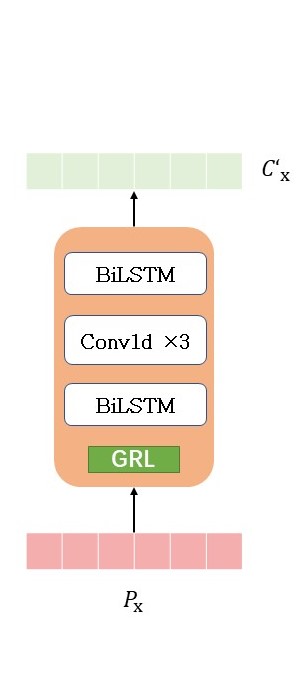}
        \end{minipage}
    }
    \caption{Architecture of PMVC. $X$ means the mel-spectrograms of the input speech. $Z_X$ are the hidden feature representations, which contain the estimated content embeddings $C_X$ and estimated prosody embeddings $P_X$. \textbf{2} and \textbf{3} with multiplication symbol \textbf{x} denote the number of InstanceNorm1d and Conv1d layers.}
    \label{Network}
\end{figure}

As depicted in Figure~\ref{Network}, the design of the feature encoder is shown in Figure~\ref{fig:3(a)}, which mainly draws on the content encoder of INVC~\cite{INVC}. Different from~\cite{INVC}, we drop the process of taking the concatenate between $X$ and hidden features as the final hidden feature $Z_X$. This ensures that $Z_X$ does not contain any timbre information. In addition, a leakyRelu function is introduced as the activation function. The architecture of the content predictor is shown in Figure~\ref{fig:3(b)}, it uses two simple BiLSTM layers and three convolution layers to predict the estimated content embeddings according to the input prosody embedding. Besides, \textbf{GRL} is positioned between the feature encoder and the content predictor. Our decoder adopts the decoder of INVC as backbone. In the training process, the speaker embedding is duplicated to match the length of prosody and content embeddings. Then we concatenate them along the channel dimension which is then used as input of the decoder to reconstruct speech.

\section{Experiments}

In this section, we conduct comparative experiments for the evaluation of the proposed model's performance on many-to-many VC and zero-shot VC tasks. As the traditional VC task, many-to-many VC means that in the inference phase select the speakers who have already appeared during training process as source and target. At the same time, zero-shot VC focuses on some more difficult tasks, in which the voice of both source speaker and target speaker are unseen during training. The audio demo is available on \url{https://largeaudiomodel.com/pmvc/}. 

\subsection{Datasets and Configurations}

\begin{table*}[htp]
  \caption{Comparison of different models in traditional VC and zero-shot vc}
  \centering
  \fontsize{8.7}{7}\selectfont
  \label{Comparison}
    \begin{tabular}{ccccccccc}
    \toprule
    \multirow{2}{*}{\textbf{Methods}}&
    \multicolumn{4}{c}{\textbf{Traditional VC}}&\multicolumn{4}{c}{\textbf{ Zero-shot VC}}\cr
    \cmidrule(lr){2-5} \cmidrule(lr){6-9}
    & MCD & MOS & TSS &PSS & MCD & MOS & TSS &PSS\cr
    \midrule
    INVC & 9.18 $\pm$ 0.34 & 2.95 $\pm$ 0.93
    & 3.12 $\pm$ 0.77 &2.74 $\pm$ 0.68 & 9.41 $\pm$ 0.58 
    & 2.75 $\pm$ 0.62 & 3.01 $\pm$ 0.89 &2.66 $\pm$ 0.74\cr
    AutoVC& 7.84 $\pm$ 0.17 & 3.24 $\pm$ 1.02
    & 3.12 $\pm$ 0.86 & 2.87 $\pm$ 0.76 & 8.06 $\pm$ 0.39 
    & 3.11 $\pm$ 0.77 & 3.06 $\pm$ 0.93  & 2.59 $\pm$ 0.87\cr
    SpeechFlow & 6.67 $\pm$ 0.29
    & 3.49 $\pm$ 0.83
    & 3.55 $\pm$ 0.69 & 3.39 $\pm$ 0.88 & 6.91 $\pm$ 0.43
   & 3.51 $\pm$ 0.92 & 3.46 $\pm$ 0.87 &3.33 $\pm$ 0.95 \cr
    \midrule
    \textbf{PMVC} &\textbf{6.06 $\pm$ 0.31} &\textbf{3.64 $\pm$ 0.90}
    &\textbf{3.91 $\pm$ 0.72} &\textbf{3.58 $\pm$ 0.84} &\textbf{5.98 $\pm$ 0.44} 
    &\textbf{3.58 $\pm$ 0.73} &\textbf{3.85 $\pm$ 0.81} &\textbf{3.42 $\pm$ 0.77} \cr
    \bottomrule
    \end{tabular}
\end{table*}

Comparative experiments were conducted on the public corpus of AISHELL-3~\cite{aishell3}. This corpus is a large-scale dataset including 88035 recordings from 218 native Chinese mandarin speakers, about 85 hours in total.
In our experiments, all recordings have a sampling rate of 22.05kHz. We follow the same train/test partition and data preprocessing as~\cite{tang2023vqcl}. Specially, We set the frame length of all training recordings to 256. That is, for any speech segments longer than 256, we randomly select 256 frames, at the same time, for those speech segments with a length shorter than 256, we pad them with constant. Besides, we divide the speech into multiple segments, each segment is 2 frames to achieve the \textbf{Algorithm~\ref{alg:algorithm}}. 

For the training of PMVC model, we set the batch size to 16 and the num of update steps is 400k. We use the ADAM optimizer~\cite{adam}~($\beta_1=0.9$, $\beta_2=0.99$, $\varepsilon=10^{-9}$). To obtain the speaker embedding, we select 10 utterances of the same speaker and feed them into the pretrained speaker encoder and then average the resulting embedding. We set the weights in Eq.(\ref{eq:9}) as follow: $ \alpha = 0.5, \beta = 0.5$. Select AutoVC, INVC, and SpeechFlow models as baseline, following the training procedure in \cite{autovc, INVC, speechsplit}. Fairly, to get the result waveform, an pretrained Hifi-GAN~\cite{hifigan} vocoder is used to convert the output mel-spectrogram.

\subsection{Comparisons}

Both objective and subjective experiments are conducted to compare different models' performances in VC tasks. Detailly, the Mel-Cepstral Distortion~(MCD) is adopted as an objective measure of the distance between the converted voice from source speaker and the real one from the target speaker.The lower MCD score means better performance.Moreover, 13 native speakers are invited as participants (nine males and four females) to do subjective tests for the quality assessment. The Mean Opinion Score~(MOS) test needs every subject to choose a score on a scale from 1 to 5 for the naturalness of the converted speech after hearing them. 
The higher score indicates the opinion that the quality of hearing speech is better. Additionally, the Voice Similarity Score~(VSS) test includes Timbre Similarity Score~(TSS), and Prosody Similarity Score~(PSS), where groups of utterances undergo voice similarity rating on a scale from 1 to 5. Each group contains four converted utterances from INVC, AutoVC, SpeechFlow, and PMVC, respectively, along with one real utterance of target speaker as reference. 
In the VSS test, higher score indicates the higher similarity between the converted result and ground truth speech.

As illustrated in Table~\ref{Comparison}, for the traditional many-to-many VC, 4 speakers are randomly selected from the training set (2 male and 2 female) and their utterances for the evaluation of multi many-to-many VC. Then test utterances of each of the 4 speakers are converted to the other 3 speakers respectively.
This process generates a total of 4$\times$3 = 12 converted utterances each of which preserves the same linguistic information of speech from the 4 speakers speech but adopts the voice of the other 3 speakers. The MOS test results show that the converted speech from our model is of higher naturalness. The VSS test results show that our method surpasses INVC, AutoVC and SpeechSplit in learning better timbre and prosodic features for the converted speech, leading to an improvement in the overall conversion effect. Results of the objective and subjective tests demonstrate that compared with other baseline models, our PMVC performs better than other baseline models in VC tasks.

For the evaluation of zero-shot conversion, we select a few unseen speakers as the source and target speakers. To obtain their timbre embedding, take 10 utterances of the source and target speaker as the trained speaker encoder input, separately. As shown in Table~\ref{Comparison}, even on zero-shot condition, the proposed method still outperforms the baselines during naturalness evaluation. Moreover, compared to the synthesized results generated from baseline models, many people have the opinion that the converted results generated from our model sound more similar to the ground truth target, demonstrating PMVC’s efficiency in zero-shot VC.

\subsection{Ablation Experiments}

In this section, we first design an ablation experiment to observe the effect of $\mathcal{L}_\text{adv}$ on our framework. Specifically, we retrained our model without $\mathcal{L}_\text{adv}$ which we called 'PMVCs'. According to our assumption, without the constrain of the adversarial Mask-Predicted loss function, some content information may leak into the estimated prosody embeddings, and our model will eventually failed in the VC task. To test this hypothesis, we can leverage the trained content predictor in our model. Specifically, we randomly select 30 speeches (15 speeches selected from the training set, another 15 speeches belong to some unseen speaker.) as the input to get the estimated content embeddings and prosody embeddings of PMVC and PMVCs respectively. At the same time, based on these prosody embeddings, the pretrained content predictor will predict the estimated content embeddings. Obviously, the more accurate the prediction result is, the more overlapping information is contained in the estimated prosody embeddings and content embeddings.

\begin{table}[htbp]
   \centering
   \caption{Results of the ablation experiments.} 
    \label{table:1}
    \begin{tabular}{l c c}
     \hline
     Method & Error (traditional-VC) & Error (zero-shot VC)\\
     \hline
     PMVC & 0.76 $\pm$ 0.10 & 0.79 $\pm$ 0.14 \\
     PMVCs & 0.14 $\pm$ 0.09 & 0.15 $\pm$ 0.11\\
     \hline 
    \end{tabular}
\end{table}

The results summarized in Table~\ref{table:1} show that without $\mathcal{L}_\text{adv}$, the content predictor can easily output the content embeddings from the prosody embeddings, which indicates that the prosody embeddings contains almost all the information contained in the estimated content embeddings. At the same time, with the constrain of $\mathcal{L}_\text{adv}$, it will be difficult for the content predictor accurately predict the content embeddings from given prosody embeddings. In other words, our PMVC has a better performance than PMVCs in separating the content information and the prosodic information. 

\begin{figure}[ht]
    
  \begin{minipage}[b]{0.8\linewidth}
            \centering
            \includegraphics[width=0.92\textwidth]{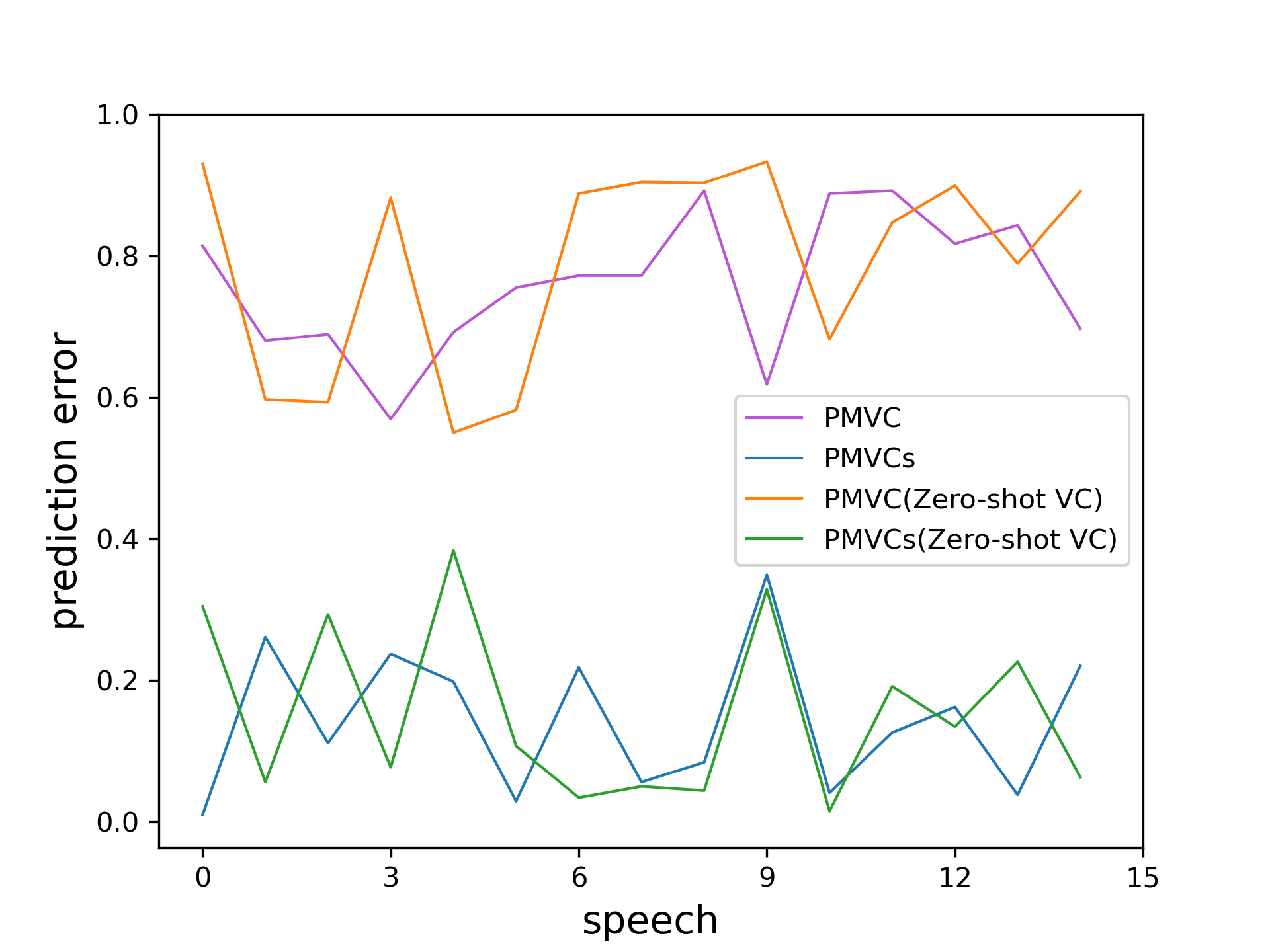}
        \end{minipage}
    \caption{Prediction errors of PMVC and PMVCs on many-to-many VC tasks and zero-shot VC tasks.}
    \label{predicterror}
\end{figure}

As visual results shown in Figure~\ref{predicterror}, the prediction error scores show that under both many-to-many and zero-shot conditions, PMVC performs better than PMVCs in decoupling prosody and content information with higher scores. In addition, we can also find that the performance of PMVC is comparable under both conditions of one-shot VC and many-to-many VC, which indicates that our model can adapt well to new unseen speakers.

In addition, to further test the above hypothesis, we prepared the ground truth speech from the source speaker and target speaker respectively. We let the subjects listen to eight converted speeches produced by our model and the retrained model respectively. If the content information of a converted speech is recognized to belong to the target speaker, it indicates a successful VC task. Otherwise, the VC task is considered to have failed.

\begin{figure}[ht]
    
  \begin{minipage}[b]{0.9\linewidth}
            \centering
            \includegraphics[width=1.0\textwidth]{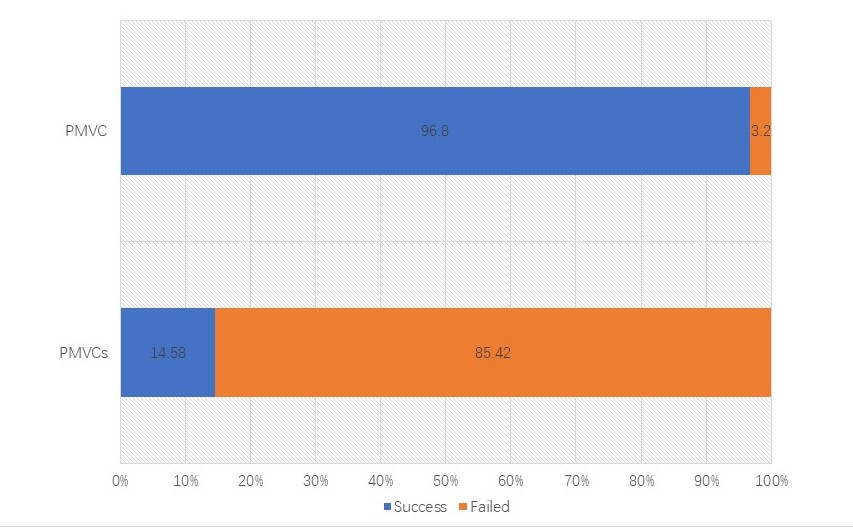}
        \end{minipage}
    \caption{Subjective evaluation for the ablation experiments.}
    \label{subjective-conversion}
\end{figure}


Results of the subjective evaluation (Figure~\ref{subjective-conversion}) indicate that almost all subjects think the performance of the retrained model in VC task is very poor. At the same time, almost all subjects believe that the proposed PMVC have achieved the VC tasks. It indicates that the adversarial loss function is crucial for the proposed framework.

Furthermore, we also verified the feasibility of using only one encoder to extract the content embeddings and prosody embeddings by designing another set of comparative experiments. Specifically, we train a new model named PMVC\_t, which mainly draws on PMVC, the only difference is that we add an additional prosody encoder to extract prosodic information in speech. And, the network structure design of the prosody encoder is almost exactly the same as our encoder. To regulate the dimension of the output feature, we simply add an linear layer at the end. To comprehensively compare PMVC and PMVC\_t, we compare their performance and inference efficiency on the VC task, respectively. 

Apart from MCD test mentioned above, we also add fake detection tests as another objective experiment, in which an open-source speech detection toolkit, \textit{Resemblyzer} (\url{https://github.com/resemble-ai/Resemblyzer}) is utilized to compare how similar 7 unknown speeches to the ground truth reference audio(6 real ones, 2 fakes which are generated from PMVC and PMVC\_t respectively).
The converted speeches are divided into 20 groups for this test. Each group contains two converted utterances generated from PMVC and PMVC\_t, respectively. The toolkit automatically assigns a score on a scale from 0 to 1 for each converted speech compared to reference audio which is ground truth. Higher score indicates that the converted speech has greater similarity to the target voice.

\begin{table}[hbp]
   \centering
   \caption{Results of the ablation experiments.} 
    \label{table:2}
    \begin{tabular}{l c c c}
     \hline
     Method & MCD Score & Detection Score &Model Size\\
     \hline
     PMVC & 5.98 $\pm$ 0.44 & 0.73 $\pm$ 0.08 & 20.8M\\
     PMVC\_t & 5.93 $\pm$ 0.35 & 0.70 $\pm$ 0.11&  31.6M\\
     \hline 
    \end{tabular}
\end{table}

As illustrated in Table~\ref{table:2}, the results show that PMVC performs on par with PMVC\_t on VC tasks. Specially, the MCD test result of PMVC\_t is slightly better than that of PMVC, and the detection score of PMVC is slightly better than that of PMVC\_t. However, the model size of PMVC is much smaller than PMVC\_t, which means smaller parameters and faster training speed. All the above comparison results show that the proposed PMVC only using one encoder to extract content information and prosodic information can significantly improve the efficiency without reducing the quality of the converted speeches.

\subsection{Flexible Hidden Features Dimensions}
\begin{figure}[ht]
    \vspace{-1.4em}
    \centering
    \subfigure[Our model]{
        \label{32-256}
        \begin{minipage}[b]{0.475\linewidth}
    \includegraphics[width=1\textwidth]{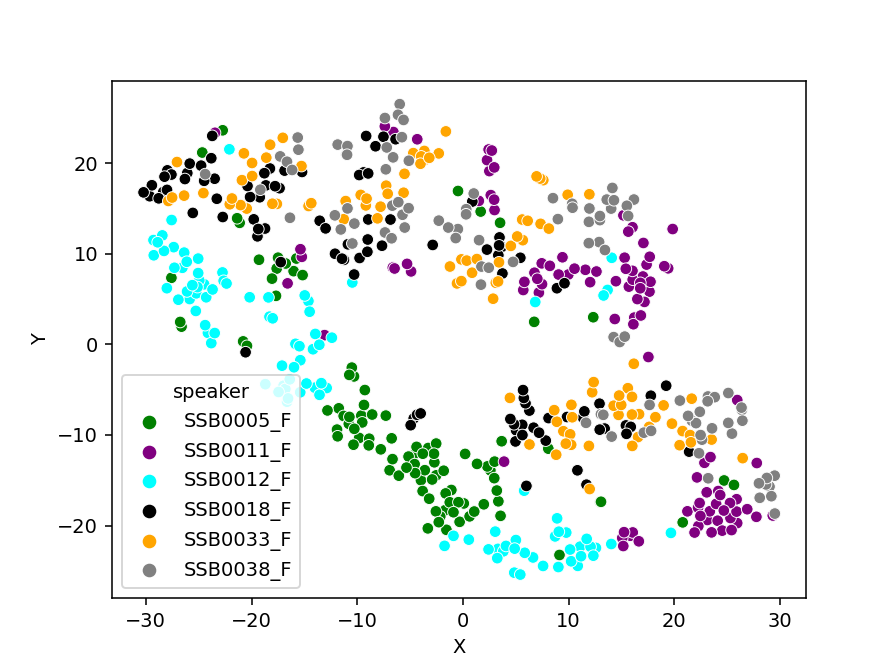}
        \end{minipage}
    }
    \subfigure[M1]{
         \label{32-64}
        \begin{minipage}[b]{0.475\linewidth}
            \includegraphics[width=1\textwidth]{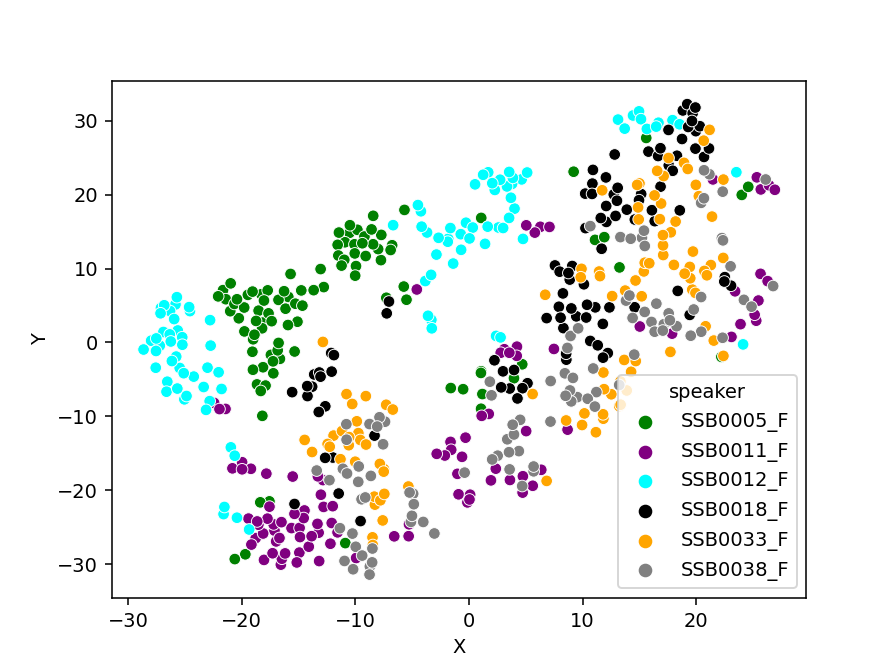}
        \end{minipage}
    }
    \subfigure[M2]{
        \label{64-64}
        \begin{minipage}[b]{0.475\linewidth}
            \includegraphics[width=1\textwidth]{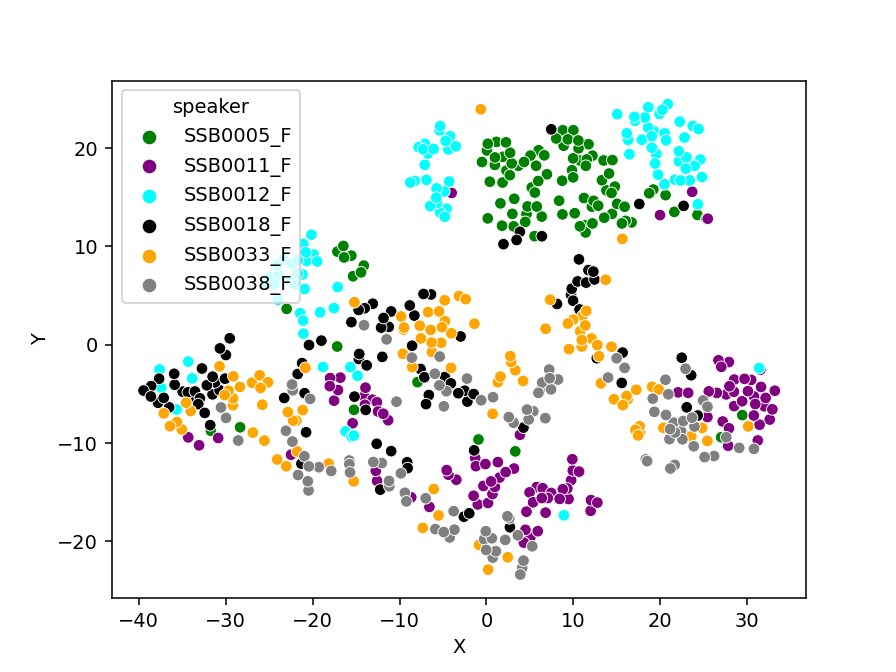}
        \end{minipage}
    }
    \subfigure[M3]{
        \label{64-32}
        \begin{minipage}[b]{0.475\linewidth}
        \includegraphics[width=1\textwidth]{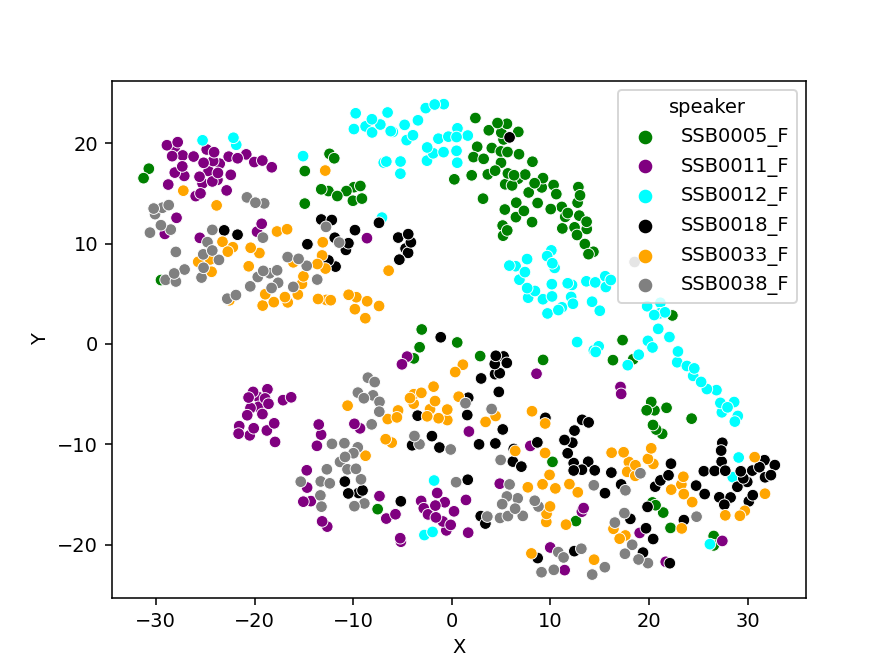}
        \end{minipage}
    }
    \caption{The visualization of hidden features. }
    \label{dim}
    \vspace{-1.5em}
\end{figure}
In this section, we will discuss the strategy to divide the latent space $\boldsymbol{Z}$ into the content embedding $\boldsymbol{C_x}$ and prosody embedding $\boldsymbol{P_x}$. Selecting the right bottleneck size is crucial in AutoVC and SpeechSplit to preserve content information while excluding timbre details.
But in our model, as we mentioned before, with the constrain of AdaIN and the loss function $\mathcal{L}_{\text {sim}}$ and $ \mathcal{L}_{\text {recon}}$, even if we do not strictly limit the dimension of the feature embeddings, the $\boldsymbol{Z}$ tends to be split into two parts representing content and prosody information ideally.
This enables us to easily determine the channel dimensions of $\boldsymbol{C_x}$ and $\boldsymbol{P_x}$ allowing for convenient extraction of both content embedding and prosody embedding using a single encoder. 

To verify that the equivalent performance of the proposed model configured with different partition modes, we retrain the proposed model by modifying the length of dimensions of $\boldsymbol{C_x}$ and $\boldsymbol{P_x}$. Specifically, in the original model PMVC, both $\boldsymbol{C_x}$ and $\boldsymbol{P_x}$ have a channel-dimension of 128. Then, the model is retrained by changing their dimensions to 96 and 160, named 'M1', or, to 64 and 192, named 'M2'. Also, we trained other models 'M3' and 'M4', which are set symmetric to 'M1' and 'M2'. Specifically, in 'M3', the dimensions of content and prosody embeddings are 160 and 96, and their dimensions change to 192 and 64 in 'M4'. Input the selected speakers' utterances (100 utterances for each) to these models and derive the estimated hidden features $\boldsymbol{Z}$~($\boldsymbol{C_x \oplus P_x}$), which then we visualize in 2D space using t-distributed stochastic neighbor embedding~(t-SNE). As illustrated in Figure~\ref{dim}, it's noted that the content and timbre information have evident separation regardless of the division proportion in the latent space.

 The performance of these models under different partition modes were assessed in the VC task by exploiting the fake speech detection toolkit \textit{Resemblyzer} again. Different from the above, in this time, each group contains five converted speeches generated from M1, M2, M3, M4 and our model, respectively. We present the result in Table~\ref{table:4}.

\begin{table}[htbp]
   \centering
   \caption{Comparison with retrained methods in VC tasks.} 
    \label{table:4}
    \begin{tabular}{l c}
     \hline
     Method & Detection Score \\
     \hline
     PMVC ($\boldsymbol{C}$ : 128, $\boldsymbol{P}$ : 128)&0.74 $\pm$ 0.12 \\
     M1 ($\boldsymbol{C}$ : 96, $\boldsymbol{P}$ : 160)&0.72 $\pm$ 0.09 \\
     M2 ($\boldsymbol{C}$ : 64, $\boldsymbol{P}$ : 192)&0.75 $\pm$ 0.11 \\
     M3 ($\boldsymbol{C}$ : 160, $\boldsymbol{P}$ : 90)&0.74 $\pm$ 0.13 \\
     M4 ($\boldsymbol{C}$ : 192, $\boldsymbol{P}$ : 64)&0.72 $\pm$ 0.07 \\
     \hline 
    \end{tabular}
\end{table}

From Table~\ref{table:4}, the results indicate that even under the changing division allocation of latent space, our model have equivalent performance in VC tasks. M2's score seems to be slightly higher than others. We attribute this to the higher channel dimensions of the prosody embedding, enabling finer modeling of prosody details, which might influence the model's VC performance.

Furthermore, we also try some subjective experiments for evaluation in VC task. In practice, 13 human participants are invited to hear a real speech and four converted speeches produced by our model and the retrained models respectively. They need to evaluate the similarity and select the converted speech which achieve the most similarity to the ground truth. Additionally, if it is difficult to judge, they can also choose the 'Fair' option.

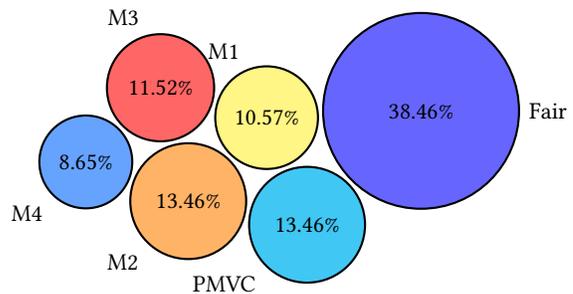
\begin{figure}[ht]
    \vspace{2em}
    \centering
    \begin{minipage}[b]{0.9\linewidth}
          \begin{tikzpicture}[scale = 0.7]
            \pie[cloud, rotate = 180]
            {38.46/Fair, 13.46/PMVC, 10.57/M1, 13.46/M2, 11.52/M3, 8.65/M4} 
        \end{tikzpicture}
    \end{minipage}
   \caption{Subjective comparison of the converted speech.}
    \label{Reconstruction speech}
\end{figure}

Results shown in Figure~\ref{Reconstruction speech} indicate that in VC task, the retrained model performs slightly worse than our proposed model, which further supports our hypothesis. That is, the performance of the proposed framework is compatible with different division modes.

\section{Conclusion}
In this paper, we propose a novel framework to address the problem of prosody modeling for expressive voice conversion. We firstly design a new random prosody algorithm to destroy the prosodic information of the source speech and obtain the corresponding augmented speech. Then, we extract and model the content, timbre, and prosodic features by using information-theory guided approaches. Both the subjective and objective experimental results demonstrate that the proposed method has made an improvement in both quality of the synthesized speech and improves its similarity to the target voice in VC tasks.
\section{Acknowledgement}
This paper is supported by the Key Research and Development Program of Guangdong Province under grant No.2021B0101400003. Corresponding author is Jianzong Wang from Ping An Technology (Shenzhen) Co., Ltd (jzwang@188.com).
\clearpage

\balance
\bibliographystyle{ACM-Reference-Format}
\bibliography{aaai22}
\end{document}